# Conference Summary: Magellanic System – Stars, Gas and Galaxies

J. Bland-Hawthorn (School of Physics, University of Sydney)
J.S. Gallagher (Dept. of Astronomy, University of Wisconsin)

**Abstract**

We provide a brief overview of some key issues that came out of the IAU 256 symposium on the Magellanic System (http://www.astro.keele.ac.uk/iaus256).

## 1. Introduction

We would like to start by thanking Jacco van Loon and the science organizing committee for proposing such a timely meeting, and for locating it in this serene corner of England. We suspect that only a tiny fraction of Britain is aware that Keele is a one-pub rural village in the heart of Staffordshire. The meeting was well organized and the facilities were excellent.

Meetings on the Magellanic System seem to come round once a decade, with the last held in 1998 in Canada. Over the past week, we have been treated to an extraordinary range of new observations over the Magellanic Clouds (MCs). The meeting has been well structured around putting the MCs into a cosmological context. The key questions that we were asked to address were:

> How does metal abundance influence star formation and stellar feedback?
> How does the structure of the multiphase ISM depend on the host?
> How has the interaction between the Galaxy and the MCs shaped their evolution?

After hearing a week's worth of lectures, we suggest, post facto, a couple of other questions that come to mind.

> To what extent have the *baryons* in the MCs evolved along different tracks relative to the Galaxy? To what extent are the MCs typical of sub-$L_*$ galaxies?

A feature of this meeting has been the prevalence of detailed wide-field maps of the Magellanic System across almost the entire electromagnetic system. These are at comparable resolution and high sensitivity to the extent that we can consider a bolometric approach at each pixel over a very wide field. This enabled many of the speakers to show their specific area of study in a wider context, an approach that was highly effective. After all, we are studying entire galaxies that happen to be on our doorstep so why not take advantage of this fact wherever possible.

To achieve the detailed comparisons that we witnessed, several groups have organized themselves into large teams (e.g. SAGE: PI Margaret Meixner) that carries with it interesting sociological issues discussed below. It also is not at all clear that the theoretical community is ready for the data deluge that is almost upon us.

As expected, Spitzer Space Telescope (SST) infrared observations featured in many talks. IRAC observations, particularly at 6-8 microns, have successfully picked up PAH emission in metal-depleted gas within the MCs. Very small grains are prominent in the MIPS 24 micron band, and large grains in the longest MIPS bands. The MIPS bands are particularly sensitive to dust in high-starlight intensity environments, e.g. star forming regions. PAHs have yet to be detected in the Magellanic Bridge and Stream where metallicities are down to 1/10 solar; this may be consistent with the very low PAH abundance expected at these metallicities (e.g. Draine et al 2007).

Several groups presented detailed star formation histories across the entire MC system. Just a decade ago, there were few YSOs known in the MC. Since Spitzer, we now see YSO populations across the entire disk of the LMC, while HST also is revealing hoards of pre-main sequence stars in young regions. It is encouraging the extent to which the multi-polyhedral CMD approach (q.v. Coimbra proceedings) is now widely used to analyze resolved stellar populations, and that different groups arrive at the same star formation histories. The stellar populations and gas/dust properties of the MC appear to be quite distinct from the Galaxy. This is evidenced by a range of different stellar components in the MC: a distinctive compact source (BH, XRB+Be) population, distributions of variable star types, and s-process chemical signatures in the RGB population, to name a few.

There is now clear evidence that stellar evolution is directly affected by the mean metallicity of the ISM. As is well known, there are distinct blue and red globular cluster populations, where the former are not seen in the Galaxy. Star cluster mass functions look distinct from the Galaxy in the sense that "infant mortality" seems much less prevalent. Moreover, a few of these get up to $10^4$ stars, which seems extraordinary in some respects.

Some of us were left reeling at the news that the HST ACS proper motions (Kallivayalil et al 2006) are now confirmed with a third epoch of observations, and by the analysis of an independent group (Piatek et al 2008). The extraordinarily high orbital speed of the MC system (~380 km s$^{-1}$), a 50% increase on what was believed just a few years ago, was not anticipated by the plethora of published dynamical models since the 1970s. Several enterprising researchers have already exploited these results, but the "standard" orbit families will have to be re-addressed in light of this.

Interesting developments are the high baryon to dark matter ratio in both MC systems, and the possibility that these galaxies entered the outer Galaxy as a group. With the promise of future astrometric missions, we will soon be moving from kinematics to dynamics for much of the Local Group. This will provide tighter constraints for ΛCDM simulations that start from an inversion of the local density field, which routinely produce better galaxies than full-blown simulations.

## 2. Major developments since 1998

We start by listing what strike us as major developments since the last conference on the Magellanic System:

- All Local Group dwarfs contain ancient stars, including MCs
- MCs have distinct *baryons* (gas+stars) compared to the Galaxy
- MC star clusters survive long term relative to our Galaxy
- The Galaxy is *not* made up of today's dwarfs
- Remarkable change in orbit parameters for MC system
- MCs have very extended stellar distributions
- Growing complexity of dark matter for MC system?
- MCs possibly accreted as a group?
- MCs have high baryon/dark matter ratio: stripped outer halos?
- Stream, Bridge metal poor relative to MCs, remnants of large gas disks?
- Stream dissolving into Galactic halo
- Distinct magnetic field structure in MCs

We have stopped short of adding hundreds of references to the above list in order to achieve some measure of brevity for the proceedings. The interested reader can quickly track back from the body of contributions here, or contact us. Some of these lead naturally into "future questions" that we address below.

We were struck by the compelling case that was made by van der Marel & Cioni (2001) for the importance of angular projection effects in the LMC in particular. The offset bar looks much more central and the outer disk is found to be intrinsically elliptic once the 3D projection effects are taken into account.

A major aspect of the meeting is the extent to which the MCs have become profoundly important astrophysical laboratories, an issue that we pick up in the next section. Gary Da Costa asked the question in reverse: where has there been essentially no progress since the 1998 meeting. One topic that comes to mind perhaps is the thorniest of them all – what are the processes involved in massive star formation? But this is clearly one of the key research areas and we would be surprised if fundamental progress had not been made in this arena by the end of the next decade.

## 3. The Magellanic Clouds as astrophysical laboratories

While observing during the early 1830s from the Cape of Good Hope, Herschel resolved the MCs into myriads of stars, thereby laying the first foundations for extragalactic astronomy. This also set the stage for the MCs to serve as laboratories where key astrophysical processes could be readily observed over a wide range of scales. At this meeting, the tradition continues into a new century with the added advantage of increasingly high angular resolution, sensitivity and panchromatic coverage across the spectrum.

With the advent of FIR surveys from the SST and AKARI, connections between the ISM and stellar populations in the MC are becoming clearer. In this way we are extending our reach into the details of the central baryon lifecycle of IGM-ISM-stars-ISM-IGM-…, which are connected by radiation and gastrophysical processes that define much of the present day visible universe. Improving data on MC chemical abundances, also tracers of this cycle, reveal differences from the Milky Way. Some of these can be traced to the varying star formation histories, but some hint at surprisingly long term ISM-IGM interactions in the MCs. Our increasingly detailed snapshots of the MC ISM further emphasize the dynamic state of this system. Stars are pumping dust and new elements into the ISM, which in turn is modified by shocks, flows, and the stellar radiation fields. A qualitative modeling of these increasingly well-defined actions stands as a key step where MC studies contribute towards a full theory of galaxy evolution.

The traditional role of the MCs as test beds for ideas concerning stellar physics remains at full value. While the HST and SST combine to explore populations of young and dying stars, wide area surveys provide full samples and reveal rare and unexpected features of stellar populations. For example, the SMC seems to prefer Be X-ray binaries and eschews black hole systems. With their weak tidal fields, low stellar densities, and favorable viewing angles, the MC are a field of dreams for star cluster studies. They are taking on a leading role in understanding the birth and survival of star clusters as well their evolution as astrophysical systems that test our ability to model real n-body systems.

In recent years, the pendulum has swung in the direction of emphasizing differences between the MCs and the Galaxy. But one should not lose sight of the fact that, given that such differences are well established now, if there are populations that are essentially identical between both systems, this is of profound importance in its own right. While there are aspects of stellar astrophysics that are largely independent of their metallicity (cf. Russell-Vogt theorem), the same goes for gas/dust phases in both galaxies.

The ways in which the MCs respond to their multiply felt external perturbations from each other and the Milky Way offer increasingly sharp tests for models of galaxy dynamics. In the MCs, we have an interacting system of galaxies that is increasingly well mapped in phase space as the quality of distance and proper motions improve. That their behaviour doesn't quite match our current expectations (e.g. structure of the Magellanic stream; stellar tidal debris) is cause for optimism that the MCs are yielding insights into hierarchical galaxy evolution. And doing so in a situation where microlensing surveys strongly constrain the nature and distribution of low luminosity baryonic matter.

**4. Near field vs. far field: what are we learning?**

We have a great deal to learn about the processes of galaxy formation and evolution from our nearest neighbours. Dwarf galaxies continue to challenge ΛCDM N-body simulators at all cosmic epochs. Are younger versions of the MCs quite typical of high redshift gas-rich galaxies? Are MCs typical of high-redshift GRB hosts? It is tempting to think so, but the quantitative case merits further development.

It seems plausible that the MCs were once part of a group that accreted onto the Local Group, and indeed such groups are being found now in the Local Universe (Tully et al 2006). Thus, it is likely that dwarf galaxies were clustered in the early universe. It is interesting to speculate to what extent that the MC accretion event onto the Galaxy is typical of group accretion in general. We may ultimately find with detailed dynamical models that we can account for one or more of the other Galactic companions or streams as progenitor group companions.

At the present time, the most detailed baryonic simulations on the scale of individual galaxies come from semi-analytic models. Arguably, they have limited predictive power and are more of a consistency check. For example, no published simulations foresaw the extent to which the baryonic content (e.g. black hole / stellar population, gas / dust content) of the MCs evolved along a distinct track from the Galaxy.

The formation and evolution of the MCs must be complicated. The high baryon/dark matter ratio suggests that the once larger dark-matter halos have been tidally stripped, at least in part. Their evolution today is further complicated by their motion through the outer Galactic halo ("halos within halos"). The system is clearly losing gas to the Galactic halo, but somehow at least the LMC appears to have ongoing gas accretion.

A long held assumption is that Galactic accretion involves baryonic matter associated with dark matter. Indeed, the dark matter stabilizes the gas against disruption by the halo corona, at least for a while. This view is partly supported by the existence of the gas-rich MCs within 50 kpc of the Galaxy.

There is possible evidence for ram-pressure induced star formation along the leading edge of the LMC. For now, we can only speculate on galaxy-wide feedback processes that operate in a dynamic pair such as this. Are there large-scale galactic winds operating in either or both systems? These would have to compete with the inevitable cross wind due to the motion of the MCs through the halo. What are the expected kinematic signatures in such a configuration? Are we already seeing these signatures along QSO sight lines?

## 5. Present and future facilities

Here are just some of the present and future facilities that have the potential to make an impact on Magellanic System research. We have removed northern facilities like PanSTARRS that are not ideally situated for this field and cannot claim completeness; so this can be thought of as a minimal summary:

> HST (serviced), VLTi, SST(warm), AKARI (warm)
> VLT, Magellan, SALT, Gemini + Adaptive optics (MCAO, MOAO, ...)
> VISTA, VST, SkyMapper

> Chandra, XMM, Suzaku
> INTEGRAL, GLAST, Herschel, Planck
> WFMOS, HERMES, …
> ALMA, ASKAP, LOFAR, NANTEN2
> JWST, GAIA, SIM, JDEM (SNAP, ADEPT, …)
> ELTs (GMT, TMT, E-ELT)
> SKA

The world has invested in an amazing array of astronomical capabilities. Many of these can and should be applied to MC studies as part of our continued development of a quantitative understanding of the universe. The lively discussions and range of ideas presented here in Keele is the basis for optimism that we will progress along this path as the opportunities arise. The one point of concern is whether the resources will exist for theory to keep pace with the observations?

**6. The sociology of large science teams**

The MCs are complex systems whose properties are accessible in detail. Large data sets and associated sophisticated models in turn require diverse expertise to obtain full scientific value. While we empathize with White's (2007) lament on the trend of astronomers to organize themselves into oversized teams, the specialized requirements of multi-disciplinary research forces this upon many of us. We can expect more work in this vein; e.g. from the upcoming major ground-based surveys and new capabilities such as Herschel in space. The challenge then is for us to be aware of the concerns associated with team science while continuing to make the best of this approach.

MC research has the advantage of relatively well-defined goals (for astronomers!) that largely follow from the standard approach of comparing observations to models which in turn are based on theory. As we've seen and heard here, we now have the further advantage of working with wide and deep seas of data, which require the efforts of many people to produce, let alone analyze. At issue is our ability to maintain focus on key scientific goals without being overwhelmed by the many details, and to creatively engage a wide skill set in the process. Organizing our intellectual resources is of increasing importance in 21$^{st}$ century MC astronomy.

Large collaborations also raise a number of well-known practical issues, especially in times of tight funding. We seem to have landed in an era that is extraordinarily rich in facilities but perhaps somewhat less so in supporting the individual scientists. Obviously we want to make the best of this situation, and to a significant degree this will mean working in teams. Care needs to be taken to maintain a healthy level of competition; history is not kind to single behemoth models of human enterprise. Giving credit where credit is due also can become difficult, especially in the context of standard astronomical practices with regard to authorship on refereed papers, an issue that is receiving wide discussion in the bio-sciences and has led to very formal procedures in high energy physics. Wide access to data sets, especially in digested form, as well as results from modelling efforts can spread the possibility for creative work beyond the individual teams. Our MC programs with their clearly prescribed data sets and model parameters can benefit from and should support effective and wide access to information, e.g. through the Virtual Observatory effort.   We've done quite well so far, and should plan to sustain this success into the future.

**7. Relevant aside: how are we to approach complexity?**

This is a fundamental question that is rarely discussed in polite company. After all, we are scientists and our job is to construct a well posed question (essentially a null hypothesis $H_o$ with a control sample) that allows for the possibility of a rejection of the hypothesis. But this question has certainly unsettled more than a few applied fields of science in recent years. John Horgan, in his provocative book "The End of Science," suggests that the era of reductionism is over and that the human race is facing fundamental barriers to the acquisition of new wisdom and knowledge. Robert Laughlin declares that "the central task of theoretical physics today is no longer to write down the ultimate equations but rather to catalogue and understand [complex] behaviour in its many guises…"

With regard to complex behaviour in astronomy, there are sociological issues to get over. At an Aspen meeting three decades ago, Ed Salpeter was confronted by a surly participant who quipped "but surely, that's just weather." "Aha!" he replied "someone who understands weather. Can you please stand and explain it to us?" The reader may be surprised to know that models for the phenomenon of rain have normalizing constants that differ by many orders of magnitude! And yet nobody would question the existence of rain, least of all a person from Staffordshire.

There is a wonderful quote by von Neumann from a speech he gave in Montreal in 1945: "Many branches of both pure and applied mathematics are in great need of computing instruments to break the present stalemate created by the failure of the purely analytical approach to non-linear problems." In fact, today, we realize that all fields of applied science (and mathematics for that matter) degenerate quickly into a wash of complexity. So now we are confronted by the prospect of acquiring vast amounts of information without receiving new wisdom. How are we to proceed?

Interestingly, we routinely make progress when there is a strong motivation to do so, e.g. medical science, environmental science, corporate finance. In most instances, we are teasing out weak signatures that are fundamental to the process at hand.

Lev Landau had an extraordinary capacity to tame complex problems. To paraphrase, consider a problem that is parametrized in the following way:

$$f = f(u_1, u_2, u_3, u_4, ...)$$

In other words, we write down all conceivable variables $u_1$, $u_2$, … that might describe the process at hand. We leave nothing out. In the pre-supercomputer era, one is forced then to search long and hard for limiting and transitional cases at the boundaries of the polyhedral space that might be amenable to attack. Only along the interstices can you even consider constructing $H_o$ because here only a limited number of variables are operating. Once the interstices and vertices are defined (often with key discoveries along the way), we can now identify a large volume in *f*-space where one can certainly extract information, but little or no physical insight. One can think of this as finding the boundaries of a problem in our search for higher organizing principles.

Nowadays, we resort to massive computer calculations and physical algorithms with ever increasing sophistication. The *f*-space is then searched exhaustively. In this way, extraordinary progress has been made in understanding turbulent media under quite specific conditions.

We cannot help feeling that the gastrophysics/ISM community needs to embrace the large suite of mathematical tools that are available today. After all, cosmologists do this as a matter of course to extract physical parameters from the microwave background, and to compare numerical simulations with large galaxy surveys. This is precisely what the gastrophysics community needs to do in order to compare the multiphase density and velocity maps which already are available for the MCs with their complex simulations. It is well known that physical processes manifest themselves in the power spectrum of the density structure, even in the presence of 3D projection effects. For example, these mathematical tools are the mainstay of the many hundreds of Los Alamos PhDs that have studied complex fluid dynamics in the presence of an extreme impulsive event.

## 8. Where will we be in 10 years?

Predictions in any field are notoriously unreliable, but for what it's worth, here goes. Even while Moore's law shows evidence of breaking down, a moderately secure prediction is that computers will be considerably more powerful, and no doubt our gastrophysical and stellar atmospheric algorithms will continue to improve. Similarly we can hope to approach more closely to a complete theoretical description of stars, including their rotation and tendency for binarity, and from this build a more accurate set of evolutionary models.

We share the enthusiasm of this conference for the impending revolution that is likely to be brought on by improved astrometry, first with the HST servicing mission providing even longer timelines on the proper motion of the MCs. Astrometry will undergo a revolution in an era of GAIA and SIM in the next decade. We will steadily move from kinematics (geometry of motion) to 3D orbit and internal dynamics, all in full 3D perspective throughout the Milky Way system.

One can envisage detailed and accurate simulations of dynamical friction, oval distortion, tidal shocking and disruption between the two galaxies, triggering star formation bursts that are consistent with the stellar record. It is not at all clear just how well this will work out. On the one hand, one of both of the MC may be accreting gas in a stochastic manner. But the new hydro simulations of the dissolving Stream, presented at this meeting, also suggest a replenishment rate for the Stream (i.e. a mass loss rate from the MC system) of around 0.1-0.3 $M_\odot$ yr$^{-1}$.

We keenly await a time when a convergence of near-field and far-field cosmology is achieved, i.e. cradle to grave observations across cosmic time from the great ALMA and JWST experiments. This prospect was foreseen by Hoyle (1965) in his Princeton & Cambridge lectures:

> It is not too much to say that the understanding of why there are different kinds of galaxy, of how galaxies originate, constitutes the biggest problem in present day astronomy. The properties of individual stars that make up the galaxies form the classical study of astrophysics, while the phenomena of galaxy formation touches on cosmology. In fact, the study of galaxies forms a bridge between conventional astronomy and astrophysics on the one hand, and cosmology on the other.

Stellar astrophysics is widely touted as one of the central pillars of modern astrophysics, but a vast amount remains to be understood. After all, just a few years ago, M.A. Asplund announced that the Sun does not in fact have canonical solar abundances (well, [Fe/H] needed to be downgraded by a hefty 0.2 dex). We know too little about non-LTE processes and accurate stellar ages continue to elude us, even with the best efforts of asteroseismologists. We were entranced by the prospect of the Eddington satellite measuring the mean molecular weight of the core of 100,000 dwarfs, but this seems to have gone the way of the dodo. Self-detonating supernova and stellar wind models seem a remote prospect but are clearly essential to future progress. Is it too much to expect major breakthroughs here?

We strongly encourage would-be theorists to give gastrophysics a serious look. In just a few years, we will be deluged with a staggering array of multiphase maps of gas, dust, particles and molecules spanning a wide range of ionization states and transitions. The stopgap divisions among ISM dust particles of aromatic features, very small and big grains looks set to subdivide further. This burgeoning field is crying out for an army of talented young recruits. Several groups around the world already treat multiphase turbulent plasmas, either with a view to studying the diffuse warm or hot media in galaxy clusters. The close association of $H_2$ and $H^+$ within cluster gas and in the M82 wind reminds us of the importance of pressure in driving the formation of dust and molecules, even in the most unlikely environments. In recent years, a few teams have made impressive headway with studying the synergistic relationship of $H_2$ and HI, clearly an important step on the road to understanding the dynamical state of the interstellar medium. From here, it's "onward ho!" to a self-consistent model of star formation, with magnetic fields in tow. Good luck!

Who knows what awaits us with the commissioning of the Large Hadron Collider later this year. The speed with which the community accepted a non-zero value of $\Lambda$ a decade ago was astonishing after two independent groups announced their high-redshift supernova results. Maybe the Dark Sector will invade our thinking again in just a few years with equal rapidity – time will tell.

## 9. Big future questions

It is abundantly clear to us that research on the Magellanic System is an extraordinarily rich area of study, at least for Southern observers. Here we have the opportunity to study two rather typical Magellanic-type galaxies in their entirety, their mutual interaction, and their interaction with the Galaxy. Dwarf galaxies continue to challenge $\Lambda$CDM predictions in the non-linear regime at all observable epochs. Here are some of the big questions we foresee in the next decade:

- How did the MCs form and how different were they from what we see today? Are there pre-ionization fossils within?
- Were the MCs much bigger at an earlier epoch? If so, where are the missing baryons and dark matter today?
- How did the MCs subsequently evolve? How is this affected by halos in halos?
- Did the MCs enter the Galactic sphere of influence as a group? Are there other identifiable dwarfs or streams that date back to this event?
- How far do their baryons and DM halos extend today?
- Do we have a complete baryon inventory?
- Where are the accreting baryons demanded by the recent SFH of the LMC?
- How did the bar form in the LMC?
- Do the MCs sustain large-scale winds?
- What are the eventual fate of the Bridge and Stream?
- How much can we learn of baryon evolution from MCs?
- How much can we learn of black hole evolution from MCs? Are there massive black holes ($>10^2$ $M_o$) lurking there today?
- How did their magnetic fields seed and evolve?

## 10. Epilogue

At this meeting, we were graced by the presence of Mike Feast who has witnessed a few astronomical revolutions in his time – after all, his first Nature paper was published 60 years ago! Mike commented that the early days of MC research were very exciting because new things kept turning up all the time and there was no real framework to build upon. But today, he senses that we are arriving at something of a synthesis, i.e. a coherent picture is emerging at least to the baryon content of both galaxies.

It was said more than once at this meeting that the MC system may be a somewhat unusual accretion event since it is difficult to find a counterpart in the nearby universe. In fact, we would disagree with this point of view. In particular, the M81 group is dominated by a large galaxy and at least three dwarfs caught up in a maelstrom of HI gas. To our view, this does not look so different from the Galaxy system. Both M31 and the Galaxy appear to have their own gas streams which are clearly much more extensive at low column density than current observations may suggest.

It is worth noting that the Local Group, dominated by the two great rajahs encircled by courtiers, is in fact quite typical of the Universe today, e.g. as demonstrated by Tully's or

Karaschentsev's group catalogues of the Local Volume. There is nothing "pathological" about the Local Group, the MC system or the galaxies within.

Let us conclude by saying that the pace of development in MC research appears to be quickening to the extent that we may not want to wait another decade to meet again. The conference chair has suggested that the next meeting be held in the Southern Hemisphere, with the faintest suggestion of South Africa as a fitting venue.

Heartfelt thanks to the Organizers of this IAU Symposium both for an excellent menu of science that was matched by the scenic venue. JSG thanks NASA for support of his Magellanic Cloud interests through funding for several HST GO and IDT research programs, and the University of Wisconsin-Madison for its investments in research support. JBH is supported by a Federation Fellowship from the Australian Research Council.